\input{psfig.sty}
\documentstyle[12pt]{article}
\newcommand{\journal}[4]{{\em #1~}#2\,(19#3)\,#4;}

\newcommand{\pr}{\journal {Phys. Rev.}}

\newcommand{\jmp}{\journal {J. Math. Phys.}}
\newcommand{\rmp}{\journal {Rev. Mod. Phys.}}
\newcommand{\cmp}{\journal {Comm. Math. Phys.}}

\newcommand{\np}{\journal {Nucl. Phys.}}
\newcommand{\pl}{\journal {Phys. Lett.}}

\newcommand{\prep}{\journal {Phys. Reports}}
\newcommand{\ptp}{\journal {Progr. Theor. Phys.}}

\makeatletter
\@addtoreset{equation}{section}
\makeatother

\newcommand{\G}{\Gamma}

\newcommand{\f}{\phi}
\newcommand{\g}{\gamma}
\newcommand{\k}{\kappa}

\newcommand{\mn}{{\mu\nu}}

\renewcommand{\o}{\omega} \renewcommand{\O}{\Omega}

\newcommand{\s}{\sigma} \renewcommand{\S}{\Sigma}
\renewcommand{\t}{\theta}

\newcommand{\EE}{{\cal E}}
\newcommand{\FF}{{\cal F}}
\newcommand{\GG}{{\cal G}}

\newcommand{\LL}{{\cal L}}
\newcommand{\MM}{{\cal M}}

\newcommand{\OO}{{\cal O}}

\newcommand{\RR}{{\cal R}}

\newcommand{\VV}{{\cal V}}

\newcommand{\ZZ}{{\cal Z}}

\newcommand{\complex}{{\kern .1em {\raise .47ex
\hbox {$\scriptscriptstyle |$}}
    \kern -.4em {\rm C}}}
\newcommand{\real}{{{\rm I} \kern -.19em {\rm R}}}
\newcommand{\rational}{{\kern .1em {\raise .47ex
\hbox{$\scripscriptstyle |$}}
    \kern -.35em {\rm Q}}}
\renewcommand{\natural}{{\vrule height 1.6ex width
.05em depth 0ex \kern -.35em {\rm N}}}
\newcommand{\integers}{{\bf Z}}
\newcommand{\ipr}{\!\cdot\!}
\newcommand{\xint}{\int d \! x \, }

\newcommand{\tr}{{\rm {Tr} \,}}
\newcommand{\Det}{{\rm {Det} \,}}
\renewcommand{\exp}{{\rm \ {exp}\,}}
\newcommand{\cb}{{\bar c}}
\newcommand{\ob}{{\bar \o}}

\newcommand{\ah}{{\hat\alpha}}
\newcommand{\gh}{{\hat g}}
\newcommand{\half}{\frac 1 2}
\newcommand{\quarter}{\frac 1 4}
\newcommand{\pa}{\partial}
\newcommand{\pad}[2]{{\frac{\partial #1}{\partial #2}}}
\newcommand{\fud}[2]  {{\displaystyle{\frac{\delta #1}{\delta #2}}}}

\newcommand{\sla}{\raise.15ex\hbox{$/$}\kern -.57em}

\newcommand{\twiddle}{\lower.9ex\rlap{$\kern -.1em\scriptstyle\sim$}}

\newcommand{\vev}[1]{\left\langle {#1}\right\rangle}

\newcommand{\equ}[1]{~(\ref{#1})}

\newcommand{\Vm}{{V_\MM}}
\newcommand{\eq}{\begin{equation}}
\newcommand{\eqn}[1]{\label{#1}\end{equation}}
\newcommand{\eea}{\end{eqnarray}}
\newcommand{\eqa}{\begin{eqnarray}}
\newcommand{\eqan}[1]{\label{#1}\end{eqnarray}}
\newcommand{\ba}[1]{\begin{equation}\begin{array}{#1}}
\newcommand{\ea}[1]{\end{array}\label{#1}\end{equation}}
\newcommand{\eqac}{\begin{equation}\begin{array}{rcl}}
\newcommand{\eqacn}[1]{\end{array}\label{#1}\end{equation}}

\renewcommand{\pad}[2]{{\displaystyle{\frac{\partial #1}{\partial #2}}}}

\newcommand{\intx}{\int d^4 \! x \, }

\newcommand{\one}{{\bf 1}}
\renewcommand{\sb}{{\bar \s}}
\newcommand{\gb}{{\bar \g}}

\begin{document}
\def\ftoday{{\sl  \number\day \space\ifcase\month 
\or Janvier\or F\'evrier\or Mars\or avril\or Mai
\or Juin\or Juillet\or Ao\^ut\or Septembre\or Octobre
\or Novembre \or D\'ecembre\fi
\space  \number\year}}    
\titlepage
%
{
\begin{center}

{ \huge    Gauge Group TQFT\\ \vskip .3ex  and \\ \vskip 1ex 
Improved Perturbative Yang-Mills\\ \vskip .3ex Theory}
\vspace{2ex}

{\Large Laurent Baulieu\footnote{e-mail~:
Baulieu@lpthe.jussieu.fr}}\\{\it\large LPTHE~\footnote{ Boite 126, Tour 16, 
1$^{\it er}$ \'etage,
        4 place Jussieu,
        F-75252 Paris CEDEX 05, FRANCE}\\ Universit\'e Pierre et Marie
Curie - PARIS VI\\ Universit\'e Denis Diderot - Paris VII\\
Laboratoire associ\'e No. 280 au CNRS }\\{\Large \vskip.5ex and \\ \vskip.5ex
Martin Schaden\footnote{e-mail~: schaden@mafalda.physics.nyu.edu\\
\indent{~~Research} supported in part by the National Science Foundation
under grant no.~PHY93-18781}}\\ 
{\it\large Physics Department, New York University,\\ 4 Washington Place,
 New York, N.Y. 10003}
\end{center}
\vspace{4ex}

\begin{center}
\bf ABSTRACT
\end{center}
{We reinterpret the Faddeev-Popov gauge-fixing procedure of
Yang-Mills theories as the definition of a topological quantum field
theory for gauge group elements depending on a background connection.
This has the advantage of relating topological gauge-fixing ambiguities to
the global breaking of a supersymmetry. The
global zero modes of the Faddeev-Popov ghosts  are handled in the
context of an equivariant cohomology without breaking translational
invariance. The gauge-fixing involves
constant fields which play the role of moduli and modify the behavior
of Green functions at subasymptotic scales. At the one loop
level physical implications from these power corrections are 
gauge invariant. }

PACS: 11.15.-q 11.15.Bt 11.15.Tk 11.15.Bx\hfill\break
NYU--TH--96/01/05\hfil HEP-TH/9601039\hfil January 1996
\vfill

\newpage
\def\be{\begin{eqnarray}}
\def\ee{\end{eqnarray}}
\def\nn{\nonumber}
\section{Introduction} In this paper we interpret  the Faddeev-Popov
gauge fixing method of Yang-Mills theories as the construction of a
Topological Quantum Field theory (TQFT) for  the gauge group.  The
gauge group element, the Faddeev-Popov ghost and anti-ghost together
with the  Nakanishi-Lautrup Lagrange multiplier field form
the BRST quartet   counting for zero degrees of freedom.  We first
define a topological  BRST invariant partition function   in the
gauge group   depending on  a background Yang-Mills connection and
thereafter average  over background connections with a gauge
invariant weight. We thus obtain a partition function that does not
dependent on arbitrary local redefinitions of gauge group elements.

The manipulations in this construction are formally  the same as
those of the Faddeev-Popov procedure.  However, the requirement of defining a
consistent TQFT for the gauge group elements immediately suggests
that one should investigate the possible breaking of the
supersymmetry.  This will permit us to consider the   Gribov
question~\cite{gr78}  
from a new point of view.   We shall see that there are obstructions
to the definition of the TQFT due to the existence of constant zero
modes for the Faddeev-Popov operator $\pa\cdot D^A$.
These zero modes occur for all transverse Yang-Mills connections $A$. If the
theory is defined on a finite space-time volume, the constant ghost modes are
normalizable and cannot be ignored. Unless they are consistently
gauge fixed, they in fact lead to a vanishing partition function in
covariant gauges. We remove these zero modes by introducing constant
ghosts and ghosts 
of ghosts in the context of an equivariant cohomology.  
In contrast to pointed gauges~\cite{si78} (where the ghost
fields are fixed at a particular space-time point), our procedure
is covariant and avoids breaking the translation invariance of
space-time. We will not treat gauge-field dependent
zero modes of the Faddeev-Popov operator in this paper. 

We can however show that the mean values of some 
observables of the gauge group TQFT     depend on global
topological properties of the background connection $A$ and we
construct a BRST-exact observable with ghost number zero whose
expectation value does not vanish on  certain manifolds.  The BRST
symmetry is found to be (globally) broken 
for finite manifolds with topological properties which also imply a
(topological) Gribov ambiguity~\cite{gr78,si78,vb95}.  This is an explicit
verification of Fujikawa's previous conjecture~\cite{fu83} that the
BRST-symmetry could be broken as a consequence of the Gribov
ambiguity. 

There is apriori no reason why the 
infinite volume limit should not be a smooth one, and we therefore
believe that the constant modes should also be gauge-fixed in this
limit. It should be stressed that our equivariant construction
neither  destroys the usual perturbative scheme of Yang-Mills 
theory nor its renormalisability. The constant ghosts however generate 
nonlocal interactions which can lead to power corrections in the usual
perturbative Green functions if certain (bosonic) ghosts have vacuum
expectation values in the infinite volume limit. The power corrections
only become important as one goes away from the asymptotic kinematical domain
where the usual Faddeev-Popov prescription is justified. We compute
these corrections for the transverse gluon propagator
perturbatively to one loop in
generalized covariant gauges. At this level, they are
found to be gauge independent provided one suitably identifies 
expectation values of the constant (bosonic) ghosts. 

\section{TQFT in the gauge group}
Consider the Euclidean Yang-Mills  partition function in Landau gauge obtained 
by the usual Faddeev-Popov construction 
 \eqac
Z[j]&=&\int [dA] \Det(\pa\ipr D^A) \delta(\pa\ipr A) \exp
[\xint ( \LL_{cl}(A) + j\ipr A)] \\&&\\
&=&\int [dA][db][dc][d\cb]\exp[ \xint (\LL_{cl}(A) +\cb
\pa\ipr D^A c - b\pa\ipr A + j\ipr A)]  
\eqacn{genfunction}
Here $A$ is the Yang-Mills connection and $D^A$ the associated
covariant derivative. 
$\LL_{cl}$ is a gauge invariant Lagrange density, for instance the
local curvature, $\LL_{cl}=\quarter\tr (F^2_\mn (A))$. In the last
expression of\equ{genfunction} the delta function and  nonlocal
determinant  in the measure 
are represented by functional integrals over the Lagrange multiplier
$b(x)$ and the Grassmannian Faddeev-Popov ghost $c(x)$ and $\cb(x)$.

The partition function\equ{genfunction} suffers from the Gribov
ambiguity~\cite{gr78}  and our aim
is to define a TQFT that leads to an improved generating 
functional which is protected from the generic zero modes of the
operator $\pa\ipr D$. 
The field variables of this TQFT are the gauge group elements $U(x)$,
which in the $SU(n)$ case are unitary matrices
depending on the space-time position $x$. The gauge transformation of
a connection $A(x)=A_\mu(x)dx^\mu$, a one-form valued
in the Lie-Algebra   $\GG$ of the gauge group $G$, is
\eq
A^U(x) =U^{-1}(x) A(x) U(x) + U^{-1}(x) d U(x) 
\eqn{gaugetrafo} 

A gauge theory is  a quantum field theory  which does not distinguish
between representations of the field variables by the configuration 
$\{A(x)\}$ or $\{A^U(x)\}$.  In  the language of TQFT,   it is a
theory  defined by a Lagrangian density   depending  locally on   $U$
in  such a way that the expectation value of certain "observables"
does not depend on  arbitrary  local  redefinitions of the $U $'s.
Inspired by our present understanding of TQFT's~\cite{bi91}, we thus
tentatively consider  the following topological   BRST symmetry
\be 
\label{top}
sU(x) &=& \Psi(x) \nn \\
s\Psi(x) &=& 0 \nn \\
s\bar\Psi(x)&=&b(x)\nn \\
sb(x)&=&0 
\ee
The fields  $\Psi(x)$ and $\bar\Psi(x)$ are topological anticommuting 
ghosts and antighosts for $U(x)$, and 
$b(x)$ is a Lagrange multiplier  field. 

Since $U(x)$ is  a group element, it has an inverse,
 and     we can introduce the  $ \GG$-valued ghost
$c(x)$ by
\be\label{defc} 
\Psi(x)=U(x)c(x)
\ee
 With this change of variables, one has
\be 
\label{base}
sU(x)&=&U(x)c(x)\nn\\
sc(x)&=& -\half[c(x),c(x)]\nn\\
s\cb(x)&=&b(x)\nn\\
sb(x)&=&0
\ee 
We have redefined $\bar\Psi$ into $\cb$ for the sake of notational
consistency. One obviously has $s^2=0$. For the particular case of
unitary groups, which we will exclusively discuss in the following,
the relation 
$UU^{\dagger} =1$ becomes consistent with  the
$s$-operation provided  $c^\dagger=-c$. The
definition of the hermitian conjugate ghost $c^\dagger$ is thus independent
of $U$. The change of variable  expressed by\equ{defc} is also  conceptually
interesting, since it provides  us with an intrinsic geometric
definition of the Faddeev-Popov ghost, with
\be
\label{c}
 c(x)= U^{-1}(x) sU(x)
\ee
 
We introduce the gauge field
configuration $ A(x) $ as a background, that is 
\eq
sA(x)=0
\eqn{brstA}
Using the definition\equ{gaugetrafo}, one finds  
\eq
s A^U(x)= -D^{A^U}c(x)\ 
\eqn{brstAU}
where 
\eq
D^A   = d +[A(x),\ ] 
\eqn{covd}
is the   covariant derivative.

Let us define the TQFT for the fields
$\{U(x)\}$ by a BRST-exact local action of the form
\eq
S_{A }[U,c,\cb,b] =\xint sW (A^U, c,\cb, b)
\eqn{action0}
and the corresponding partition function
\eq
\ZZ [A]=\int [dU][dc][d\cb][db]\ \exp S_{A }  
\eqn{gentqft}
$W$ in\equ{action0} is a local functional of the fields $c,  \cb,
b$ and  $A$  with total ghost number~$-1$. By expanding $s W$ as
a function  
of all fields, one obtains  an action which is $s$-invariant, since
$s^2=0$. The resulting BRST symmetry   can be interpreted as a
twisted version of supersymmetry~\cite{bi91}. If this supersymmetry
is unbroken, the usual arguments based on the definitions of bosonic
and fermionic path integration would show that $\ZZ [A]$ is a sum of ratios
of  determinants, such that
\eq
\ZZ [A]\neq 0\qquad {\rm and} \qquad \fud{\ZZ [A]}{A}  =0
\eqn{unbroken}

Consider  the following choice for the topological action,
\be\label{gaugeaction}
S_A &=&- 2\xint  s \tr \left(\cb(x)  \pa\ipr A^U(x)\right)\nn\\
  &=& 2\xint \tr \left (\cb(x)\pa\ipr D^{A^U} c(x) - b \pa\ipr A^U\right ) 
\ee
The expectation value of a gauge invariant field
functional $\OO[A]$  is defined as
\be\label{genalpha}
\vev{\OO[A]}   \propto  \int [dA]   \OO[A]\exp [ \xint
\LL_{cl}   ]\  
 \int [dU][dc][d\cb][db]
\exp  S_A
\nn\\
\ee  
Suppose now that the BRST symmetry is
unbroken. Then\equ{unbroken}  is true and\equ{genalpha} factorizes
into the expectation value of $\OO(A)$  with respect to the gauge invariant
measure $[dA] \exp  \xint \LL_{cl}  $ and a nonvanishing
  normalization. Gauge invariance of this  measure and the observable
$\OO$  furthermore implies that the change of variables 
\eq
A\rightarrow A^\prime =U^\dagger A U +U^\dagger d U\  
\eqn{chvar}
decouples  the integration over the volume of the gauge-group   and one has
\be\label{genFP}
\vev{\OO[A]}  \propto\int [d{A'}][dc][d\cb]  \OO[{A'}]\exp  \xint \left( 
\LL_{cl}({A'}) + 2\tr(   \cb \pa\ipr D^{A'}  c - b{ \pa\ipr
{A'}})\right ) \nn\\ 
\ee

We have thus reproduced  the Faddeev-Popov  construction  by coupling a
supposedly trivial TQFT in the gauge group to the gauge-invariant Yang-Mills
theory. We have   used   the gauge-invariance of the measure $[dA]$,  of
the classical action $\xint
\LL_{cl}$ and of the observable $\OO$ to factorize the integration
over the gauge-group.

It is clear that all defects of the Faddeev-Popov construction must
be present in this derivation, which  at first sight  is only a
change of terminology. Indeed, the definition\equ{genfunction} of
$Z[j]$   only  makes sense perturbatively  since the condition
$\pa\ipr A=0$ does not select unambiguously 
the representative $A$ of the gauge field~\cite{gr78,si78}. This
is due to the existence of   disconnected gauge
transformations which imply a   nontrivial moduli-space for the
equation $\pa\ipr  A=0$.  Thus,  the statement\equ{unbroken} must
be wrong in our derivation,
and the point we wish to make is that  the Gribov   ambiguity
is related to a breaking of the supersymmetry of the topological
action\equ{gaugeaction}.

The origin of the SUSY-breaking is quite clear:   the operator
$\pa\ipr D^{A}$
has zero modes. Thus the quadratic form $\xint \cb(\pa\ipr D^{A})c$ is
degenerate and there is a deficit in the supersymmetric compensations
which  would otherwise enforce\equ{unbroken}. Having $\ZZ [A]=0$ on
a domain of the connection $A$ with nonzero measure would destroy
the meaning of the Faddeev-Popov construction.

We can distinguish two types  of ghost and antighost zero modes, the
constant ones and those which are space-time  dependent. The
nonconstant zero modes generally depend quite strongly on the
connection $A$. The latter type of zero modes can  be avoided in the
formulation of 
the TQFT by choosing an appropriate background $A$. They  could
however be important in the YM-theory, if the set of connections
$A$ with such zero modes has non-vanishing 
measure (see ref.~\cite{zw94}).  Generally however, their existence
does not seem to jeopardize the definition of the Yang Mills path
integral itself~\cite{hi79,le95} although they could render a
perturbative evaluation of it untractable~\cite{zw94}. 

We will focus on  the constant zero modes, since they are the only
ones which are an obstruction to the definition of the
TQFT\equ{gentqft}, for {\it any} background connection $A$. These
zero modes are present for all transverse connections $A$ (which
certainly is a set of nonvanishing measure). Due to the fact that,
$\partial\ipr D^A= D^A\ipr \partial$, for transverse connections,
the YM-action in Landau gauge has  the obvious on-shell symmetry, 
\be\label{mode}
c(x)\to c(x)+const.\nn\\
\cb(x)\to \cb(x)+const.
\ee
The associated zero modes are normalizable if the base-manifold is
compact and imply that eq. (\ref{unbroken}) does not hold for
any $A$ in this case. The above symmetry\equ{mode} is a
consequence of the rigid gauge invariance of the covariant gauge
fixing and the partition function\equ{gentqft} of the TQFT with the 
action\equ{gaugeaction} is proportional to the Euler number $\chi(
SU(n) ) =0 $ of the global $SU(n)$ group manifold of fixed
points~\cite{bi91}. This
destroys the possibility to reach the Faddeev-Popov action, since the
partition function of the gauge group TQFT vanishes. For finite
space-time volume the
symmetry\equ{mode} therefore has to be removed and we believe this is 
also necessary to define a smooth infinite volume limit. 

 The situation is  
analogous to that  in string theory where  one cannot   globally gauge-fix
the 2-dimensional metric to a background metric. The field theory
signals   this obstruction  by  the existence of zero modes for the
reparametrization antighosts. The remedy is well-known. It
consists in weakening  the over gauge-fixing  by introducing 
  finite integrations over  constant moduli, genus by
genus~\cite{fr82}. This procedure    
can be interpreted as a BRST invariant gauge fixing term for the global zero
modes of the Faddeev-Popov operator of string theory~\cite{ba88}.

The  analogy with  string theory suggests  that one
should also  gauge-fix in a BRST-invariant way the degeneracy of the
Yang-Mills  action with respect to constant translations of the ghosts and
antighosts.  The method we will use is covariant and fixes the global
symmetry while preserving invariance under space-time translations
(unlike pointed gauges~\cite{si78,mi81}). It is then possible to
consistently define the   
partition function of the gauge-fixed Yang-Mills theory by averaging
over the (background) connection of the gauge group TQFT in a gauge
invariant way.   This gauge-fixing therefore commutes with
space-time symmetries which after all is the main reason for
considering covariant gauges and their associated ghosts.   
 
\section{Gauge-fixing the global zero modes}
To define the Yang-Mills theory, we   must therefore include in the
BRST algebra the symmetry     (\ref{mode}). The theory will
involve  a system of
$\GG$-valued global ghosts associated with the constant
modes of the Faddeev-Popov ghosts.  The geometrically meaningful sector of
the BRST symmetry has positive ghost number. To isolate
consistently the gauge-fixing of the constant ghost zero modes from the usual
Yang-Mills gauge-fixing, we    consider the following equivariant
BRST symmetry  modulo constant gauge transformations 
\be\label{basicbrs}
s U(x) &=& U(x)(c(x)+\o)\nn\\
s c(x) &=& -\half [c (x), c (x)] - [\o, c(x)] -\f\nn\\
s \o &=& -\half [\o , \o ] +\f\nn\\
s \f &=& -[\o, \f] 
\ee
where $\o$ and $\f$ are $x$-independent.
 The commuting $\GG$-valued  constant ghost of ghost  $\f$   
corresponds to  the anticommuting parameter
 of the translational symmetry
for  the ghost $c(x)$, $ c(x)\to c(x)+constant$. Moreover the positive
degrees of freedom carried by
$\f$ compensate the  negative ones carried by the
$\GG$-valued anticommuting constant ghost $\o$ and thus the
effective number of degrees of freedom remains unchanged. In other
words, one can intuitively justify introducing an equivariant
cohomology to cure the redundancy in the replacement of $c(x)$ by 
$c(x)+\o$. 

The gauge conditions we impose are
\be\label{constraint}
\partial\ipr A^U(x)&=& 0\nn\\
\xint \ c(x) &=& 0\nn\\
\xint \ \cb(x) &=& 0
\ee
These choices preserve the translation and the Euclidean invariance
of space-time.

The topological action is obtained by implementing the
constraints\equ{constraint}  
with Lagrange multipliers $b(x)$, $\sb$ and  $\g$ 
  having ghost number~$0$, $-1$ and $1$ respectively. One   can extend 
the action of $s$ to these multiplier fields without destroying the
nilpotency  by introducing two more $\GG$-valued constant ghost
fields $\gb$ and $ \s$. 
The canonical dimensions and ghost numbers of the fields are  summarized in
Table~1
\begin{center}
\begin{tabular}{|l|r||r|r|r|r|r|r|r|r|r|r|}\hline
field&$A(x)$ & $U(x)$ & $c(x)$ & $\cb(x)$ & $b(x)$ & $\f$ &
$\o$ & $\s$ & $\sb$ & $\gb$ & $\g$\\ \hline
dim&$1$&$0$&$0$&$2$&$2$&$0$&$0$&$4$&$4$&$2$&$2$\\ \hline
$\f\Pi$&$0$&$0$&$1$&$-1$&$0$&$2$&$1$&$-2$&$-1$&$0$&$1$\\ \hline
\end{tabular}

\vspace{.2cm}{\footnotesize
{\bf Table 1.} Dimensions and ghost numbers of the fields.} 
\end{center}
 and the 
action of the BRST-operator $s$ on all fields is
\be\label{sdef}
s A_\mu(x) &=&0\nn\\
s U(x) &=& U(x)\o + U(x)c(x)\nn\\
s c(x) &=& -[\o, c(x)] -\half [ c (x), c (x)] -\f\nn\\
s \o &=& -\half [ \o,\o] +\f\nn\\
s \f &=& -[\o, \f]\nn\\
s \cb(x)&=&-[\o,\cb(x)] + b(x)\nn\\
s b(x) &=& -[\o, b(x)] + [\f, \cb(x)]\nn\\
s \s &=& -[\o,\s]  +\sb \nn\\
s \sb &=& -[\o, \sb] + [\f, \s]\nn\\
s \gb &=& -[\o,\gb] +\g\nn\\
s \g &=& -[\o, \g] + [\f, \gb] 
\ee

It is straightforward to show that this BRST-operator is 
nilpotent  on any element of the graded algebra constructed from the
fields of Table~1
\eq
s^2=0
\eqn{nilpotency}
Notice that one has
\be
sA^U_\mu(x)=D^{A^U}_\mu c(x)-[\o, A^U_\mu(x)]
\ee
We thus see that  $\o$ generates constant gauge transformations for all
other fields of the BRST algebra. Moreover, the introduction of
$\o$ permits  us   to handle the invariance of 
$\pa\ipr A^U=0$ with respect to constant gauge transformations.
The BRST-exact topological action which implements the
constraints\equ{constraint} is 
\be\label{actionA}
S_A=s W_A &=&    2\int_\MM dx\ s \tr [\pa^\mu \cb(x) A^U_\mu(x) +
\gb  \cb(x) + \s  c(x)]\nn\\
&=&2\int_\MM dx \tr  \ [  \pa^\mu b(x)  A^U_\mu(x) - \pa_\mu\cb(x) 
D_\mu^{A^U} c(x)   \nn \\ 
&&\qquad  \qquad +\g \cb(x) + \gb b(x) + \sb c(x) -
\half\s  [c(x),c(x)] -\s \f\ ]  
\ee
The nilpotency of $s$ guarantees that $S_A$ is BRST  invariant. Due
to this invariance the constant modes of the fields
$b(x)$ and $\cb(x)$  are simultaneously eliminated. Note that for an
{\it abelian} group the bosonic ghosts $\s$ and $\f$
 decouple completely. 
  
$S_A$ is   independent
of the  anti-commuting  ghost $\o$, generating constant
gauge transformations. As a consequence, the partition function 
\eq
\ZZ[A] =\int [dU] [dc] [d\cb] [db] d\f d\o d\s d\sb
d\gb d\g\  e^{S_A}
\eqn{partition0} 
vanishes due to the Grassmann-integration over $\o^a$.  Notice
that the  integration over the corresponding bosonic zero modes of
constant $SU(n)$ transformations only gives a factor proportional
to the finite volume  of the (global) $SU(n)$ group manifold.
Table~1 shows  that the
measure in\equ{partition0} has net ghost-number $n^2-1$. For a nonzero
partition function $\ZZ[A]$, one has to absorb the
excess Grassmann modes by inserting a factor $\prod_{a=1}^{n^2-1}
\o^a $ in the measure. Equivalently, one can  drop the $d\o$
integration in\equ{partition0}. 

The action $S_A$ is  invariant under the transformations $s_\o$,
parametrized by $\o$
\be\label{restricts}
s_\o \o &=&0\nn\\
s_\o &\equiv& s {\rm\ \ on\ all\ fields\ except\ } \o\ 
\ee
which here plays the role of an external (Grassmann-)parameter.
$s_\o$ in general is not a nilpotent operation on all fields,
since for instance 
\eq
s_\o^2 c(x)   =[ \o^2-\f, c(x)] \neq 0\   
\eqn{breaking} 

The elimination of the constant ghost $\o$ from the partition function can
also be achieved by  a more conventional gauge-fixing, using the
additional fields defined in Table 2, 
\begin{center}
\begin{tabular}{|l|r||r|r|r|r|r|r|r|r|r|r|}\hline
field&$\alpha$ & $\ob$ & $\beta$  \\ \hline
dim&$0$&$4$&$4$\\ \hline
$\f\Pi$&$0$&$-1$&$0$ \\ \hline
\end{tabular}

\vspace{.2cm}{\footnotesize
{\bf Table 2.} Dimensions and ghost numbers of the additional  fields.} 
\end{center}
with $s$ extended by
\be
s\alpha &=& 0\nn\\
s\ob &=& \beta\nn\\
s\beta &=& 0
\ee
Including these fields in the formalism, one can add the
following BRST-exact action to $S_A$
\be
S_{GF}&=& 2\xint s\left( \ob (\alpha+\ln U(x))\right)
\nn\\
&=&2\xint \left( \beta(\alpha+\ln U(x))-\ob (\o+c(x))\right)
\ee
This action, when inserted in the path integral, allows us to
eliminate the fields $\o, \ob, \beta, \alpha$ by Gaussian
integrations which yield ratios  of equal  normalisation factors,
i.e. a factor of one.  One can thus formally justify dropping the
integration over $\o$ in\equ{partition0} at the price of
introducing a multivalued function of $U(x)$ in an intermediate step.

Whichever explanation one prefers for the absence of $\o$ in the
final action\equ{actionA}, one is forced  to only consider
expectation values of $\o$-independent BRST invariant
functionals.  These functionals are globally gauge invariant since
$\o $ generates constant gauge transformations. We therefore
define observables as elements of  the equivariant cohomology 
$\S$,
\eq
\S =\{\OO   :\pad{\OO}{\o^a}=0; s\OO=0, \OO\neq sF\}
\eqn{observables}
where $F$ is itself $\o$-independent. Notice that  the action of
$s_\o$ on $\o$-independent and globally gauge-invariant
functionals is equivalent to that of~$s$. The
definition of physical observables can be  more restrictive. For
instance one may require that such an observable also has vanishing ghost
number.

\section{Observables in the gauge group TQFT and global breaking of the BRST
symmetry}
According to the last section, the physically interesting expectation
values are 
\eq
\vev{\OO}_A =\int [dU] [dc] [d\cb] [db] d\f d\s d\sb
d\gb d\g\ \OO\  e^{S_A}   
\eqn{expect}
where $\OO$ belongs to $\S$.
In general, $\vev{\OO}_A$ is independent  of local variations of the
background connection $A$ since one has 
\eq
\fud{}{A}\vev{\OO}_A = \vev{\OO \fud{}{A} S_A}_A= \vev{\OO s \fud{}{A}
W_A}_A=-\vev{ (s\OO) \fud{}{A} W_A}_A =0,\,\forall \OO\in \S
\eqn{equivariant} 

We will show in the following that the constant fields introduced to
gauge-fix global zero modes of  the  Faddeev-Popov ghosts and
eventually construct a well-defined Yang-Mills partition function,
also imply the   existence of nonlocal  observables depending on
global  properties of $A$. The existence of such observables does 
not invalidate the local equation\equ{equivariant},  but it does
prevent us from extending\equ{equivariant} to the statement that  the
expectation value of an observable is globally independent of the  
background $A$. 

The existence of TQFT observables follows from Chern identities which
imply descent equations for elements of $\S$. It is convenient to
introduce the 1-form, 
\eq
L= U^{-1} d U 
\eqn{L}
which is a flat connection associated with the unitary field $U(x)$ 
\eq
  d L + L^2=0
\eqn{flat}
$\tr L^3$ is a globally gauge invariant d-closed but not d-exact
3-form with vanishing ghost 
number. It is the starting point of the descent chain,
\ba{rclrcl}
0 &=& - d\O_3^0 \quad\quad&\O_3^0 & =&\tr L^3/3\\   
s\O_3^0 &=& - d \O_2^1 \quad\quad &\O_2^1& =&\tr c L^2\\
s\O_2^1 &=& - d \O_1^2 \quad\quad &\O_1^2& =&\tr (c^2 -\f) L\\
s\O_1^2 &=& - d \O_0^3 \quad\quad &\O_0^3& =&\tr (c^3/3- c\f)\\
s\O_0^3 &=& ~ ~ \O_0^4 \quad\quad&\O_0^4& =&\tr \f^2\\
s\O_0^4 &=& 0\quad\quad &&&
\ea{descent1} 

The lower  and upper indices of a form $\O$ denote respectively
its form degree  and its ghost number. Note that consistency requires
that  $d\O_0^4=0 $  which is obviously true since $\f$ is a constant
field.    The integrals over  p-cycles of  each   p-form  ($1\leq
p\leq 3$) in these descent equations are observables since
they are elements of the equivariant cohomology $\S$. 

We can therefore identify  the winding number $\OO_0[U]_{\G_3}$ of
the map $U: \G_3\rightarrow SU(n)$ as an observable with ghost
number zero, 
\eq
\OO_0[U]_{\G_3}= \frac{1}{8\pi^2} \int_{\G_3} \O_3^0 =
\frac{1}{24\pi^2} \int_{\G_3} \tr L^3
\eqn{winding}

In what follows we will only consider this observable. It reflects
certain topological properties of the space-time manifold $\MM$: it
vanishes  for manifolds which do not admit any 3-cycle
$\G_3\subset \MM$ with $\pi_3(\G_3)=\integers$. The winding
number can for instance be nonzero  on  space-time manifolds with the
topology $S_3\times S_1$. On the other hand, it will vanish for the hypertorus.
 
The  other  forms in\equ{descent1} with lower degree
encode topological information even for manifolds where the winding
number is trivial. This encoding is however quite abstract since
the corresponding observables have nonzero  ghost number.  One can
moreover also construct observables with forms obtained by ascending from 
invariants of the constant ghost $\f$ with higher ghost number,
\eq
\O^{2n}_0=\tr\f^n \Rightarrow \O^{2n-1}_0=\tr c
\f^{n-1}+\dots\Rightarrow\dots 
\eqn{general}
 One of the more interesting forms obtained in this way is the
4-form with ghost number~1
\eq
\O^1_4 =\tr c L^4 
\eqn{anomaly} 
which  is related to the potential ABBJ anomaly and whose chain
terminates with $\tr \f^3$.
 
Using the winding number\equ{winding}, we  will now show that there is
a global dependence of the gauge group TQFT on the connection $A$ for
certain manifolds. Consider a space-time manifold $\MM$, which has at
least one three-dimensional cycle $\G_3\subset \MM$ with
$\pi_3(\G_3)=\integers$ and on which a unitary field $g(x)$ is
defined which maps $g :\G_3\rightarrow SU(n)$ with a given winding
number $\OO_0[g]_{\G_3}=w\neq 0$.  Using the
fact that the topological action $S_A$ of\equ{actionA} only depends
on $U$ through $A^U$, basic group properties and\equ{expect} imply
\be\label{dependence}
\vev{\OO_0 [U]_{\G_3}}_{A^g} &=&\vev{\OO_0[g^\dagger U]_{\G_3}}_A =
\vev{\OO_0[U]_{\G_3}}_A -\vev{\OO_0[g]_{\G_3}}_A
\nn\\
&=&\vev{\OO_0[U]_{\G_3}}_A -w \vev{1}_A
\ee
To derive this formula, one uses the property that the winding number of
the composition of two maps is the sum of their winding numbers. One
also assumes that the path integral measure in\equ{expect} is
gauge invariant -- which is plausible  in the absence of an ABBJ
anomaly\footnote{Note however that topological quantities such as the
winding number 
make little sense in a  lattice analog of this model. }. It is worth
emphasizing that the only aspect of the equivariant construction which
entered the derivation of\equ{dependence} is the fact that the
winding number\equ{winding} is an observable of the TQFT. 

Singer~\cite{si78} has shown that Gribov copies~\cite{gr78} of a gauge
connection have to occur for manifolds which admit maps with
nonvanishing winding number. Equation\equ{dependence} implies that 
either $\ZZ[A]$ in\equ{unbroken} depends globally on $A$ or vanishes for
such manifolds and we will see below that this amounts to a breaking
of the BRST symmetry. Precisely the same topological obstruction is
therefore the cause of a Gribov ambiguity as well as the breaking of
the BRST-symmetry. The relation\equ{dependence} therefore strongly
supports Fujikawa's~\cite{fu83} original conjecture that the two
phenomena could be related.

In Appendix~A we present topological arguments for the special case of
an $SU(2)$ gauge group that the partition function $\vev{\one}_A$ no
longer vanishes in the vicinity of $A=0$ for the equivariant TQFT
defined by\equ{expect} with action\equ{actionA}.  We now show
that equation\equ{dependence} in this case has the natural interpretation that
the equivariant BRST symmetry we are discussing is globally (or
spontaneously) broken. In view of Singer's~\cite{si78} result, this would
be an interesting way of characterizing the Gribov phenomenon.   Our
argument is quite formal however, since we consider the following nonlocal
functional
\eq
\OO_B  = \left(W_A - W_{A^g} \right) \sum_{n=1}^\infty \frac{1}{n
!} \left( S_{A^g}-S_A \right)^{n-1} \  
\eqn{Obreak}
$\OO_B$ is globally gauge invariant and independent of the
 ghost $\o$. One therefore has,
\eq
s\OO_B = -\sum_{n=1}^\infty \frac{1}{n!} \left( S_{A^g}- S_A
\right)^n = 1- \exp[ S_{A^g}-S_A ]  \  
\eqn{exact}
due to the definition\equ{actionA} and the nilpotency of $s$.  One
can thus cast\equ{dependence} in the
form 
\eq
\vev{ s\left\{ \OO_0[U]_{\G_3} \OO_B\right\}}_A = w\vev{1}_A \ 
\eqn{broken}
 
Provided the  map $g : \G_3\rightarrow SU(n)$ with  winding
number~$w$ exists, we therefore have a (nonlocal) $s$-exact
functional whose expectation  value for general background
connections $A$ does 
not vanish. This is the conventional indication for the spontaneous
breakdown of a symmetry.

The results for the construction of the Yang-Mills theory in Landau
gauge can be summarized as follows. By gauge fixing the constant zero
modes of the ghosts, one obtains a generally well defined {\it
nonvanishing} gauge group TQFT partition function which depends on 
global properties of the Yang-Mills background connection. One can in
principle now proceed to a definition of the gauge theory by
averaging over the background connection of the gauge group TQFT with a
gauge invariant weight. In the following, we will choose other gauges
for the gauge group TQFT in order to conveniently perform  explicit
computations and to check  the gauge independence to some extent.

We will see that the constant ghosts introduced by the equivariant
BRST\equ{sdef} can modify the Yang-Mills perturbation theory in an
interesting way. When they condense, they lead to power corrections
for physical correlation functions at large momenta.

\section{General covariant gauges} 
Since we are interested in practical computations to
investigate possible modifications of perturbation theory by the
constant ghosts, we will use the freedom of choosing different BRST
invariant topological actions to eliminate most of the
Lagrange-multiplier constraints of\equ{actionA} in favor of
interactions. Although stable under
renormalization,\equ{actionA} is not the most general BRST-exact
renormalizable action  one can construct with this field content.

By examining Table~1, one sees that the only local BRST exact terms
of dimension four, ghost number zero and independent of $\o$
which have been omitted in\equ{actionA} are
\eq
\alpha\  s\tr
\cb(x)b(x) =\alpha\ \tr (b^2(x) +\f [\cb(x),\cb(x)])  
\eqn{alphaterm}
and
\eq 
\beta \ s\tr [\cb (x) ,\cb (x)]c(x)  
=
\beta\ \tr (2 b(x)[\cb(x),c(x)]- \half[\cb (x) ,\cb (x)][c(x)
,c(x)]-\f [\cb (x) ,\cb (x)])  
\eqn{betaterm}
 where $\alpha$ and  $\beta$ are dimensionless gauge parameters. Note that
$\tr b^2$ alone would {\it not} be  exact with respect to the BRST symmetry
of eq.\equ{sdef} because of the constant ghosts.  
We now observe that in more general  covariant gauges where the TQFT
of Landau gauge described by\equ{actionA} is extended by the
BRST-exact terms\equ{alphaterm} and\equ{betaterm}, the
constant ghost $\f$ usually couples to $\intx [\cb (x),\cb (x)]$.

For $\alpha\neq 0$ the constraint $\pa\ipr A=0$ is softened  and
replaced by a Gaussian dependence on the longitudinal part of the
connection (in the gauge-theory longitudinal gluons only propagate in
gauges with $\alpha\neq 0$).  A nonvanishing value of the gauge
parameter $\beta$ generally   introduces a local quartic ghost
interaction, which leads to a more complicated perturbation theory.
At the special value $\beta=\alpha$  this quartic ghost interaction
vanishes after gaussian elimination of the field  $b(x)$   and the
resulting action is again bilinear in the ghosts $c(x)$ and $\cb(x)$.
Interestingly the constant $\f$-ghost also decouples from $[\cb (x)
,\cb (x)]$ at this point in the parameter space and $\f$ as well as
$\s$  can be eliminated from the action. Moreover the elimination
of $b(x)$ changes the Faddeev-Popov term $\cb(x) \pa\ipr D^A
c(x)$ into $\cb(x)
D^A\ipr\pa c(x)$.  Upon rescaling $\cb(x) \rightarrow
\cb(x)/\alpha,\ \sb\rightarrow\sb/\alpha,\
\g\rightarrow \g/\alpha$, the relevant topological action in
gauges with $\beta=\alpha$ is,  
\be\label{effactionA}
S^{(1)}_A(\alpha) &=&\frac{2}{\alpha} \int_\MM  dx \tr\left[\half(\pa\ipr
A^U(x))^2 + \cb(x)\, D^{A^U}\ipr\pa c(x)- 
\right.\nn\\  &&\left.\quad -[\cb(x),c(x)]\gb +\half\gb^2
 + \g\cb(x) + \sb c(x) \right] \nn\\
\ee
The other class of gauges where dynamical  ghosts effectively only enter
the action quadratically is when  $\beta=0$. In these gauges one has 
\be\label {effactionB}
S^{(2)}_A(\alpha) &=&\frac{2}{\alpha} \int_\MM dx \tr \left[\half(\pa\ipr
A^U(x))^2  +  \cb(x)\pa\ipr D^{A^U} c(x) -\right.\nn\\
&&\quad \left.-\half\s [c(x),c(x)]-\half\f[\cb(x),\cb(x)]-\s\f
+\g\cb(x) + \sb c(x) \right]\nn\\ 
\ee
where we have again rescaled the fields
$\cb\rightarrow \cb/\alpha$, $\s\rightarrow\s/\alpha$, 
$\sb\rightarrow \sb/\alpha$ in order to
exhibit clearly that the loop expansion is an expansion in the gauge
parameter $\alpha$. 

The models described by\equ{effactionA} respectively\equ{effactionB}
are all stable under 
renormalization, essentially because one of the ghost momenta factorizes in
the quartic ghost vertex, rendering it superficially
convergent. $\beta=\alpha$ and $\beta=0$ are therefore fixed lines in the
parameter space which intersect at the Landau gauge, ($\alpha=\beta=0$). 

 The dependence of the effective actions\equ{effactionA}
and\equ{effactionB} on the constant ghosts $\gb$, respectively $\f$ 
and $\s$,  can be eliminated in favor of nonlocal quartic ghost
interactions with zero momentum transfer. This is reminiscent of
Cooper-pairing, especially since the ghosts obey Fermi-statistics.
The analogy is perhaps particularly striking for the models described
by effective actions\equ{effactionA} if we consider half the
ghostnumber as the analog of the z-component of ``spin'': the
``Cooper''-pair in this case has total ghostnumber $0$ (i.e. $s_z=0$)
(it is anyhow a scalar under rotations, since the ghost-fields
themselves are and they are coupled to zero angular momentum). Such a
``Cooper''-pair is furthermore ``charged'' since it transforms
according to the adjoint representation of the gauge group. In
contrast to the abelian case however, two such ``Cooper''-pairs can
form a state of vanishing chromoelectric charge, and the global gauge
symmetry need not be broken if these ``Cooper''-pairs of ghosts
condense.   The physical interpretation of the breakdown of the
BRST-symmetry without breaking the global $SU(n)$ symmetry of the
model could thus be the condensation of such ``Cooper''-pairs. The
other class of actions\equ{effactionB} can be thought of as a
``Fierzed'' description of the same effect:  in this case there are
two types of charged pairs with ghost number $\pm 2$ (i.e.
corresponding to electrons coupled to $s_z=\pm 1$). If they condense
symmetrically, a similar physical picture may result. We will not
pursue this amusing analogy further, but will show that the physical
content of the  two classes of models described by\equ{effactionA}
and\equ{effactionB} is remarkably similar.

\section{Modification of ordinary perturbation theory}
In the following  we use the gauge group TQFT partition functions
defined with  $S^{(1)}_A $ or  $S^{(2)}_A $ instead of $S _A $, and
average\equ{expect} over background connections $A$ with a gauge
invariant weight as explained in section~2. The motivation for this
procedure is that it is not altogether trivial to implement 
Landau gauge in the mode expansion on a {\it finite}
manifold. It is thus desirable for practical calculations to replace
these constraints by a Feynman-type gauge-fixing term $(\pa\ipr
A)^2$. We can explore the gauge dependence with the gauge
parameter $\alpha$ and  Landau
gauge should be the limit $\alpha\rightarrow 0$ in either class of
gauges.   After the change of variables, $ A^U\rightarrow A $ the
unitary fields $U$  decouple as previously  and  the resulting 
action is   invariant  under the     BRST symmetry based on the equation
\eq
s A=-D^A c(x)-[\o, A]
\eqn{sA}  
The new feature is that the perturbative computations will be
modified by the interactions  of the dynamical fields with the
remaining constant ghosts in $S^{(1)}_A $ respectively $S^{(2)}_A $.

We are mainly concerned with the infinite volume
limit of the $SU(n)$ gauge theories described by the classical actions
\eq
S^{(1,2)}_G=\tr\int_\MM \frac{1}{4 g^2} F_{\mu\nu} F_{\mu\nu} +
\left. S^{(1,2)}_A( \alpha)\right|_{U=\one}
\eqn{actionF}
 
To regularize ultra-violet  divergences in the perturbative
expansion,  we use dimensional regularisation and renormalize
according to the minimal scheme. From now on, $D$ (without indices) denotes the
dimension of spacetime ($D\leq 4$).
 
The actions $S^{(1,2)}_G$ depend on the usual dynamical ghost fields $c(x)$
and $\cb(x)$ as well as on constant ghosts. The corresponding
Euclidean  field theory
is defined on a finite $D$-dimensional manifold, which for simplicity
will be taken to be a  symmetric  torus of extension $L$. The
integration over the   Grassmann variables $\g^a$ and $\sb^a$
removes the constant modes of $c(x)$ and $\cb(x)$ in the mode
expansion.  A perturbative treatment of the  couplings between the
dynamical ghosts and  the constant {\it bosonic} ghosts  $\gb$ or
$\s$ and $\f$   would lead to severe infrared divergences of
the perturbation series (just like a perturbative treatment of
mass-terms would).  We therefore propose to treat these bosonic fields like
moduli  on which the generating functional depends. The correlation
functions have to be integrated  over this space of moduli  with a
certain weight which in principle can be computed order by order in a
loop expansion  for  the dynamical fields. The exact weight is
not perturbatively computable, but we will see that one only has to
know (or assume) a few  expectation values of the constant ghosts to
effectively parametrize  the integration over the moduli-space in the
infinite volume limit.

The general structure of the moduli-space is  simplest for the gauge-theory
defined by the classical action $S^{(1)}_G$, since it only depends on
one constant bosonic ghost, $\gb$. If we only consider globally
$SU(n)$-invariant  expectation values, one can perform a global
$SU(n)$-transformation to diagonalize $\gb$. The moduli space in this
case is therefore the direct product of the $SU(n)$ group manifold and
an ($n-1$)-dimensional manifold described by the eigenvalues of $\gb$.

The moduli-space of the model described by the classical action
$S^{(2)}_G$ is of higher dimension and considerably more complicated
since it is described by $\f$ and $\s$. One can again factor
the $SU(n)$-group, but is left with a $n^2-1$ dimensional manifold,
which furthermore has one flat direction if ghost number is conserved.

Following the general argument that functionals in the equivariant
cohomology $\S$ can be  chosen not to depend on BRST
doublets~\cite{di77}, it is clear that physically interesting
observables with vanishing
ghost-number do not depend on the dynamical ghosts (nor on
$\f,\s,$ or $\gb$). We can
therefore integrate over the dynamical ghosts $c(x)$  and $\cb(x)$
that only appear quadratically in the effective action\equ{actionF}
without loosing any physical information. This results in an effective
action $\G$ which is a nonlocal functional of the gauge
connection $A$ and the constant bosonic ghost(s). Since we are
interested in the effects from the constant ghosts in a perturbative
evaluation of gluonic correlation functions, we expand $\G$ in orders of
the gauge field $A(x)$ as follows
\be\label{expansionA}
\G[\varphi,A(x)]&=& \Vm \G^0(\varphi) + \half\xint\int d \! y
\, A_\mu(x)^a 
\G^{2\,ab}_{\mu\nu} (x-y,\varphi) A_\nu^b(y) + \dots\nn\\
\ee 
The generic variable $\varphi$ in\equ{expansionA} stands for all
constant bosonic ghosts which have not been eliminated. The
$A$-independent term $\G^0$ in\equ{expansionA} is  just the 1-loop
effective action for the constant bosonic ghost(s). Note that we have
explicitely indicated that this term is proportional to the volume
$\Vm$ of space-time. The effective polarization $\G^2$
in\equ{expansionA} depends on the constant ghost(s) in a highly
nontrivial manner and leads to a correction of $\G^0$ at the 
2-loop level, etc. Since we cannot solve the gluonic part of the
theory, we are not able to calculate the true weight for the
integration over the moduli-space of the $\varphi$'s.  We can however
quite independently of perturbation theory assert that the
$A$-independent part of $\G$ is proportional to the volume of
space-time. In the infinite volume limit only the  maximum of
the exact effective action $\G^0(\varphi)$ is therefore relevant.
This fact allows us to effectively perform the integration over the
constant ghosts at any given finite order of the loop
expansion for the gauge field by evaluating expectation values at the
maximum of the full effective action $\G^0$.

We will apply this procedure to compute the effective gluon
polarization $\G^2$ at the one
loop-level as a function of the expectation value of the relevant
$\varphi$'s. The latter are  therefore   important parameters  in
a perturbative evaluation of the  theory and  we will show that
nontrivial expectation values of the bosonic ghosts give rise to
asymptotic power corrections in the loop expansion of correlation
functions.  Moreover, using two different classes of gauges
corresponding to the choices  $S_A^{(1)}$ and $S_A^{(2)}$ for the
gauge group TQFT, we can test the gauge independance of physical
correlation functions.

Let us first compute  $\G^0(\gb)$ of\equ{expansionA}  to one loop
for the theory described by the action 
$S^{(1)}_G$. We take space-time  to be  a symmetrical torus $L\times
L\times \dots
\times L$ of $D$ dimensions\footnote{Apart from the finite volume we
essentially  follow the procedure of ref.~\cite{co73}. Note that
$\G^0$ is the {\it negative} of the effective potential
of~\cite{co73} and we are therefore interested in its maximum.}. The 
classical contribution to $\G^0({\gb})$ is the quadratic term
$-\half\gb^a\gb^a/\alpha$, while the $\alpha$-independent next
order in the loop expansion is found  by evaluating the infinite sum
of one ghost loop diagrams with no external $A$ fields shown in Fig.~1.
\vskip .5cm
\hskip 4cm\psfig{figure=ghostloop1.ps,height=1.3in}
\nobreak\newline
{\small\baselineskip 5pt 
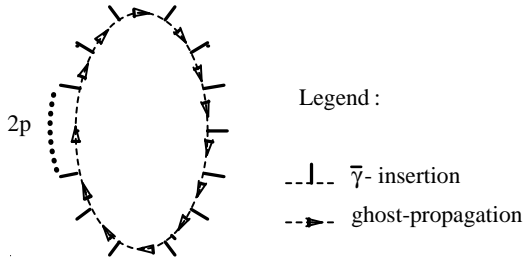
\noindent Fig.1: Nonvanishing 1-loop contribution to the effective
moduli action: a ghost loop with $2p
$ insertions of $\gb$-moduli.}  

Note that the antisymmetry of the structure constants $f^{abc}$ of a
nonabelian group implies that only loops with an even number of
$\gb$-insertions give a nonvanishing contribution to the effective action.
The  loop with $2p$  insertions of the ghost $\gb$ (which we will call a 
$p$-loop) is proportional to
\eq
C_p=\tr_{adj.} \hat\gb^{2p}
\eqn{colorfac}
where we have used the notation $\hat X$ 
to denote the
(antihermitian) matrix of the adjoint representation of $X$, $\hat
X_{ab}=f^{abc} X^c $.  
For $SU(2)$ the traces\equ{colorfac} are simply powers of the single
invariant $\gb^a\gb^a$ 
\eq
C_p(SU(2))= 2 (-\gb^a\gb^a)^p
\eqn{su2color}

The contribution from a single
$p$-loop to the effective action is
\newcommand{\sumn}{\sum_{\{n_1\dots n_D\}\neq\{0\dots 0\}}}
\eq
F_D(p)= - \frac{1}{2p}\tr_{adj.}
\left(\frac{L^2\hat\gb}{(2\pi)^2}\right)^{2p} 
\sumn \left( n_1^2 +\dots+ n_D^2\right)^{-2p}  
\eqn{sumn}
The sum in\equ{sumn} extends over all the sets of integers $\{n_1\dots
n_D\}$ describing the complete set of modes with momenta $k_\mu=2\pi
n_\mu/L$ of the dynamical ghosts. The contribution from the constant
modes with $n_\mu=0, \ \mu=1,\dots,D$,  to\equ{sumn} is however
eliminated  by the integration over the ghosts $\sb$ and $\g$.
$F_D(p)$ is therefore finite for 
$p>D/4$. The overall negative sign of\equ{sumn} is due to
the ghost statistics. 

One can analytically continue to noninteger dimensions by casting $F_D(p)$
in integral form, 
\eq
F_D(p)=  - \tr_{adj.} \left(\frac{L^2\hat\gb}{(2\pi)^2}\right)^{2p}
\frac{1}{\G(2p+1)} \int_0^\infty dx\,x^{2p-1} 
\left[\left(f(x) \right)^D -1\right] 
\eqn{FD} 
where the function $f(x)$ for $x>0$ is the convergent sum,
\eq
f(x)= \sum_{n=-\infty}^\infty e^{-x n^2} 
\eqn{fdef}
Although there is no analytic expression for $f(x)$ at arbitrary
values of its argument, the reflection formula~\cite{el89}
\be\label{reflection}
f(x)= \sqrt{ {\pi}\over{x}}\ f({  {\pi^2}\over {x}}) 
\ee
implies the asymptotic behaviour,
\be\label{asymptotic}
f(x\rightarrow\infty)&=& 1 + 2e^{-x} + O( e^{-4x})\nn\\
f(x\rightarrow 0) &=&\sqrt{\frac{\pi}{x}}(1 + O(e^{- {{\pi^2}\over x}}))\nn\\
\ee

Since the commuting ghosts $\gb^a$ can be treated as real numbers
\eq
v^2=\frac{L^4\gb^a\gb^a}{(2\pi)^4}   
\eqn{vdef}
is positive and the 1-loop contributions to the effective action  can
be  summed over all values of $p\geq 1$. Adding this sum to  the
tree-level contribution gives
\eq 
{L^D \G^0 ({\gb})}{=}{-\frac{(2\pi)^4 v^2 L^{D-4}}{2\alpha}}  {
-4\int_0^\infty \frac{dx}{x} \sin^2(vx/2) \left[\left(f(x) \right)^D
-1\right] }  
\eqn{sumE} 
The asymptotic behaviour of $f(x)$ when  $x\rightarrow 0$
determines the leading behaviour of the integral in\equ{sumE} when
$v\rightarrow\infty$. This is the large-volume limit we are
interested in. One obtains in $D=4-2\epsilon$ dimensions,
\eq
L^D\G^0(\gb)\quad \stackrel{\mu
L\rightarrow\infty}{\longrightarrow}\quad  (\mu L)^D
(-\tilde v^2)\left[\frac{(4\pi)^2 \mu^{2\epsilon}}{2\alpha}
+\frac{4}{4-2\epsilon} \tilde v^{-\epsilon}
\cos(\half\pi\epsilon)\G(\epsilon-1)\right] 
\eqn{VD}
where we have introduced the {\it finite} scale $\mu$ since we are
interested in the $\mu L\rightarrow\infty$ limit. The amplitude of 
$\gb^a\gb^a$ is now measured in terms of $\mu$ with
\eq
\tilde v^2=\frac{\gb^a\gb^a}{(4\pi)^2\mu^4}\ 
\eqn{vtilde}
The term in square brackets of\equ{VD} does not depend on the large scale
$\mu L$
and we can expand it for $\epsilon\rightarrow 0$ (but an expansion of
the factor $(L\mu)^D$ around $D=4$ would make no sense, since we want
to evaluate the effective action when $\epsilon \ln (L\mu)\gg 1$). 

The $1/\epsilon$-term in the expansion of\equ{VD} can be removed by a
counterterm of the form $(1-Z)\Vm {\gb^a\gb^a}/{2\alpha}$ in the
classical action. To order $\hbar$ we thus have in the MS-scheme,
\eq
Z=1+ \frac{2\ah \gh^2}{(4\pi)^2\epsilon}
\eqn{counter}
where we have expressed  $\alpha=\ah\gh^2\mu^{2\epsilon}$ in terms of the
dimensionless gauge-parameter $\ah$ and coupling $\gh=g \mu^{-\epsilon}$. 
The renormalization constants of the usual parameters and fields of
the gauge theories  
described by\equ{actionF} are the same as those of the 
standard Yang-Mills theory (this will be verified shortly). The
renormalization constant for the 
gauge parameter $\alpha=\ah g^2$ in the MS-scheme in the notation
of~\cite{it80} is 
\eq
Z_\alpha=Z_3 Z_g^2=1 -\frac{3\gh^2}{(4\pi)^2\epsilon} -\frac{\ah
\gh^2}{(4\pi)^2\epsilon} +O(\hbar^2)
\eqn{zalpha}
for an $SU(2)$ gauge group\footnote{$Z_\alpha$ at one loop  is 
independent of the number $N_F$ of quark flavours. This supports the
idea that the gauge fixing of global zero modes only concerns the
pure gauge boson sector}. Together with\equ{counter} $Z_\alpha$
determines the renormalization constant 
$Z_{\gb^2}$ between the bare   and renormalized    composites
$(\gb^a\gb^a)_B$   and $(\gb^a\gb^a)_R$
\eq
(\gb^a\gb^a)_B=Z_{\gb^2} (\gb^a\gb^a)_R\ 
\eqn{zdef} 
One has 
\eq
Z_{\gb^2}=Z_\alpha\,Z= 1 -\frac{3\gh^2}{(4\pi)^2\epsilon}
+\frac{\gh^2\ah}{(4\pi)^2\epsilon} + O(\hbar^2)={\tilde Z}^{-2}_3 \
\eqn{zfac}
where ${\tilde Z_3}$ is the renormalization constant for the kinetic
term $\cb\pa\ipr\pa c/\alpha$.  This result reflects the fact at the
one-loop level  that the $\cb\gb c/\alpha $-vertex in the action
$S^{(1)}_G$ is not  renormalized.

We may trade the dependence of the renormalized  one-loop effective
action on the dimensionless gauge parameter $\ah$ and scale $\mu$ for
a new scale $\k$ defined as follows
\eq
\ln \frac{\k^2}{4\pi\mu^2}=-\frac{(4\pi)^2}{2\ah \gh^2}+1-\g_E  
\eqn{scalekap}
where $\g_E$ is the Euler constant. The value
$\gb^a\gb^a=\k^4$ maximizes $\G^0$ at one loop.  In terms of
this scale, the one loop renormalized effective action $\G^0$ for
an $SU(2)$ 
gauge theory with classical action $S^{(1)}_G$ takes the simple form
\eq
L^4\G^0 (\gb^a\gb^a;\k^4)=-L^4\frac{\gb^a\gb^a}{32\pi^2 }
\ln\left[\frac{\gb^a\gb^a}{e\k^4}\right]  
\eqn{EFF}
in four space-time dimensions. 

The effective action\equ{EFF} is real and
has a unique maximum at 
\eq
\gb^a\gb^a=\k^4>0\
\eqn{expectvalue}
We assume that the exact effective action will also have a
nontrivial maximum.
Since the effective action is proportional to the volume of
space-time, the variance of $\gb^a\gb^a$  vanishes in the
thermodynamic limit and it is sufficient to know the true expectation
value $\vev{\gb^a\gb^a}$ to perturbatively evaluate any   gauge boson
correlation function in the equivariant cohomology of the $SU(2)$
theory. For a more general group than $SU(2)$, the knowledge of a
larger number of expectation values of  invariants of $\gb$ would
be  needed. This number   equals     the rank of the group.  The one
loop effective action calculated above indicates that the expectation
value $\vev{\gb^a\gb^a}$  need not vanish and furthermore correctly
predicts its anomalous dimension.  We will shortly examine the effect
 of   a nontrivial expectation value    $\vev{\gb^a\gb^a}$ on the
transverse gluon polarization. 

Let us first however address the issue of  gauge-invariance  by
also calculating the one loop effective action for the constant ghosts
$\s$ and $\f$ for the class of theories with   classical
actions  of the type $S^{(2)}_G$. The one loop  diagrams that have to
be evaluated in this case are those  of Fig.~2.   
\vskip .5cm
\hskip 4cm\psfig{figure=ghostloop.ps,height=1.3in}
\nobreak\newline
{\small\baselineskip 5pt 
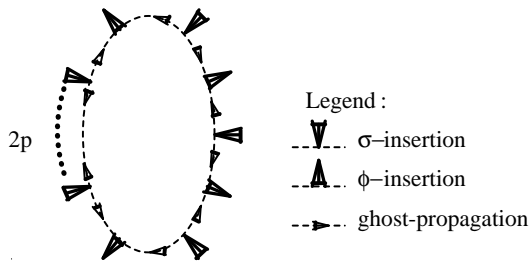
\noindent Fig.2: Nonvanishing 1-loop contribution to the effective
moduli action: a ghost loop with $p
$ alternating  insertions of $\f$ and
$\s$-moduli.}  

The structure of the interaction vertices in\equ{effactionB} requires
that a nonvanishing loop has an equal number of 
$\f$ and $\s$ insertions, which furthermore alternate. A 
calculation similar to the
previous one gives for the renormalized 1-loop effective action of the
bosonic $\s$ and $\f$ ghosts,  
\eq
L^4\G^0 (\f^a\s^a;\bar\k^4)=L^4 \frac{\s^a\f^a}{32\pi^2}
\ln\left[-\frac{\s^a\f^a}{e\bar\k^4}\right] 
\eqn{EFF2}
where the gauge-group is again $SU(2)$ and the space-time volume is large.

The unique
maximum of the effective action\equ{EFF2} occurs for
\eq
\s^a\f^a=-\bar\k^4<0 
\eqn{expectvalue2} 
and the relation between the parameter $\bar\k^4$   
and the gauge parameter $\ah$ and scale $\mu$ in
this case is
\eq
\ln \frac{\bar\k^2}{4\pi\mu^2}=-\frac{(4\pi)^2}{\ah \gh^2}
+1 -\g_E
\eqn{scalekap2}

The 1-loop effective actions\equ{EFF2} and\equ{EFF} suggest  that
$-\vev{\s^a\f^a}$   and  $\vev{\gb^a\gb^a}$ play analogous
physical roles in the gauges described by the actions $S^{(2)}_G$
and $S^{(1)}_G$. Note that the relations\equ{scalekap2}
and\equ{scalekap} determining the position of the maxima of the 
respective 1-loop effective actions in terms of the scale $\mu$ and
the gauge parameter $\ah$ differ by $\ah\rightarrow 2\ah$. A byproduct
of the calculation leading to\equ{EFF2} is 
the renormalization  constant of $(\s^a\f^a)$,
\eq
Z_{\s\f}= 1 -\frac{3\gh^2}{(4\pi)^2\epsilon} +
O(\hbar^2) 
\eqn{zfac2}
Comparing\equ{zfac2} with\equ{zfac} shows that the critical exponents of 
$(\gb^a\gb^a)$ and $\s^a\f^a$ coincide for $\alpha\rightarrow
0$. These results are consistent, since the models we are considering should
coincide at the point $\alpha=\beta=0$ in the
gauge-parameter space which corresponds to  Landau gauge.

Let us finally investigate effects on the perturbative Green
functions of the gauge boson from nonvanishing expectation values
$\vev{\gb^a\gb^a}$ and $\vev{\s^a\f^a}$.       
To this end, we  calculate the generalized gluon
polarization $\G^{2\, ab}_{\mu\nu}(x-y;\varphi)$
of\equ{expansionA} to one loop in the gauge classes defined by 
$S^{(1)}_G$ respectively  $S^{(2)}_G$. 

The dependence of the 1-loop gluon polarization on the constant
ghost $\gb$ in the model with action $S^{(1)}_G$ comes from a ghost
loop with two gluon vertices and an arbitrary number of insertions of
the constant ghost $\gb$. The only nonvanishing contributions for
$SU(2)$ correspond to the Feynman diagrams of Fig.~3.
\vskip .5cm
\hskip .2cm\psfig{figure=polarization1.ps,height=2.0in}
\nobreak\newline
{\small\baselineskip 5pt 
\noindent Fig.3: $\gb$-dependent contributions to the 1-loop gluon
polarization for $SU(2)$: a) an even number ($\ge 2$) of $\gb$-insertions on
one side of the loop only,  b) an even number ($\ge 2$) of insertions on
either side of the loop, c) an odd number of insertions on either
side. Legend as in Fig.~1.}

Summing over the number of insertions for each class of diagrams in Fig.~3
these contributions to the gluon polarization in momentum space  correspond
to the integral expressions, 
\newcommand{\intdk}{\int\!\frac{\mu^4d^Dk}{(2\pi\mu)^D}}
\newcommand{\intk}{\int\!\frac{d^4k}{(2\pi)^4}}
\be\label{contributions1}
\G^{2\,ab}_{\mu\nu}(q;a))&=&
-2(\gb^a\gb^b+\delta^{ab}\k^4)\k^{-4} \intk \frac{(k+q)_\mu 
k_\nu}{k^2 (k+q)^2}\sum_{n=1}^\infty
\left(\frac{-\k^4}{k^4}\right)^n \nn\\ 
&=& 2(\delta^{ab}\k^4+\gb^a\gb^b) \intk \frac{(k+q)_\mu
k_\nu}{k^2 (k+q)^2 (k^4 + \k^4)}\nn\\
\G^{2\,ab}_{\mu\nu}(q;b))&=&-2\gb^a\gb^b \k^4\!\intk \frac{(k+q)_\mu
k_\nu}{k^2 (k+q)^2 (k^4 + \k^4)((k+q)^4 +\k^4)}\nn\\
\G^{2\,ab}_{\mu\nu}(q;c))&=&-2\gb^a\gb^b 
\intk \frac{(k+q)_\mu k_\nu}{(k^4 + \k^4)((k+q)^4+\k^4)} \nn\\
\ee
We have     replaced $\gb^a\gb^a$  in\equ{contributions1} by its
vacuum expectation value $\k^4=\vev{\gb^a\gb^a}$ in the thermodynamic
limit. Note that the (Euclidean) integrals in\equ{contributions1} only
make sense for $\k^4>0$ and are ultraviolet and infrared finite in
$D=4$ dimensions. The fact that these corrections from the constant
ghosts are finite proves our previous assertion that the
renormalization constants of the ordinary Yang-Mills theory are
unchanged (at least in the minimal scheme and to one loop). 

A partial fraction decomposition expresses the moduli-dependent
contribution to the gluon polarization,
\be\label{partfrac}
\G^{2\,ab}_{\mu\nu}(q)=&&\k^2 \lim_{D\rightarrow
4}\left(\delta_{\mu\nu} -2 q_\mu q_\nu\frac{\pa}{\pa q^2}\right)\left[\delta^{ab}(2 I_D[q^2; 0,0]
-I_D[q^2;\k^2,0]-I_D[q^2;-\k^2,0])\right.\nn\\ 
&&{\kern +5em} +\left.\frac{\gb^a\gb^b}{\k^4}
(I_D[q^2;\k^2,0]+I_D[q^2;-\k^2,0]-2I_D[q^2;\k^2,-\k^2])\right] \nn\\ 
\ee
in terms of the dimensionless scalar integrals 
\eq
I_D[q^2;a,b] =\frac{\G(\frac{2-D}{2})}{8\pi}
 \int_0^1 dx \left[\frac{x(1-x)q^2 + i a x + i
b(1-x)}{4\pi\k^2}\right]^{\frac{D-2}{2}}  
\eqn{scalarintegral}
The limit in\equ{partfrac} exists and the leading correction
from a nonvanishing value of $\k$ to the asymptotic behaviour of the
transverse gluon polarization is 
\eq
\G^{2\, ab}_{\mu\nu}(q^2\rightarrow\infty;\gb)\propto
\frac{\k^4}{q^2} \ln q^2
\eqn{asymp}
This is not the asymptotic behaviour a
simple effective gluon mass would produce. It is remarkable
that\equ{asymp} is consistent with the leading power correction to the 
gluon propagator that   arises from a restriction to the Gribov
region~\cite{gr78} or the fundamental domain~\cite{zw94}. 

We have only been considering an unbroken $SU(n)$ gauge group so far
and one may wonder whether the constant ghosts could not also lead
to noticeable effects in the broken case. Note however that  the
reasoning leading to the equivariant construction does not
really apply to the broken case, because the $\o$ ghost
generates constant gauge transformations. An immediate consequence is
that the ghosts are generally massive in gauges which depend explicitely on the
Higgs field (for example t'Hooft gauges). The constant ghosts
$\gb,\s$ and $\f$ futhermore  
decouple if the invariance is abelian. These general arguments suggest
that one might only expect 
physical effects from the constant ghosts if an unbroken nonabelian invariance
remains. In the following we will only investigate the gauge
dependence in an unbroken 
$SU(2)$ theory  by comparing the gluon
polarization of models described by $S^{(1)}_G$ and $S^{(2)}_G$ at the
one loop level.

For the model described by the classical action $S^{(2)}_G$, the
contributions to the gluon polarization from the moduli
$\f$ and $\s$ are those shown in Fig.~4.
\vskip .5cm
\hskip .2cm\psfig{figure=polarization.ps,height=2in}
\nobreak\newline
{\small\baselineskip 5pt 
\noindent Fig.4: $\s$ and $\f$-dependent contributions to the 1-loop
gluon polarization for $SU(2)$: a) an even number ($\ge 2$) of
insertions on one side of the loop only, b) same as a) but with exchanged
gluons,  c) an even number ($\ge 2$) of insertions on
either side of the loop, d) an odd number of insertions on either
side. Legend as in Fig.~2.}

The corresponding integral expressions for these polarizations are 
\newcommand{\kb}{{\bar\k}}
\be\label{contributions2}
\G^{2\,ab}_{\mu\nu}(q;a))&=& (\s^a\f^b
-\delta^{ab}\kb^4)\kb^{-4} \intk \frac{(k+q)_\mu
k_\nu}{k^2 (k+q)^2}\sum_{n=1}^\infty
\left(\frac{-\kb^4}{k^4}\right)^n \nn\\ 
&=&(\delta^{ab}\kb^4-\s^a\f^b) \intk \frac{(k+q)_\mu
k_\nu}{k^2 (k+q)^2 (k^4 + \kb^4)}\nn\\
\G^{2\,ab}_{\mu\nu}(q;b))&=&(\delta^{ab}\kb^4-\s^b\f^a) 
\intk \frac{(k+q)_\mu
k_\nu}{k^2 (k+q)^2 ((k+q)^4 + \kb^4)}\nn\\
\G^{2\,ab}_{\mu\nu}(q;c))&=&(\delta^{ab}\kb^8\tan^2\t
-\f^a\f^b\s^2 -\s^a\s^b \f^2)
\nn\\
&&\qquad \intk \frac{(k+q)_\mu
k_\nu}{k^2 (k+q)^2 (k^4 + \kb^4)((k+q)^4 +\kb^4)}\nn\\
\G^{2\,ab}_{\mu\nu}(q;d))&=&(\s^a\f^b+\s^b\f^a) 
\intk \frac{(k+q)_\mu
(k+q)_\nu}{(k^4 + \kb^4)((k+q)^4+\kb^4)}\nn\\  
\ee
where we have replaced $\s^a\f^a$ by the expectation value 
$\kb^4=-\vev{\s^a\f^a}>0$. The dependence of\equ{contributions2} on the
opening angle $\t$,
\eq
\tan^2\t =\frac{\vec\f^2\ \vec\s^2-(\vec\f\ipr\vec\s)^2}{
(\vec\f\ipr\vec\s)^2}   
\eqn{angle}
is to be expected in this case, because it is the only other
$SU(2)$-invariant with vanishing ghost number that can be formed with
$\vec\s$ and $\vec\f$. The one loop effective action $\G^0$
of\equ{EFF2} does not depend on this invariant,
but\equ{contributions2} shows that further loop corrections 
and thus the full effective action do depend on the angle $\t$.

Note however that for the special case, 
\be\label{vacuumrel}
\vev{\tan^2\t}&=&0\nn\\
 \kb^4=-\vev{\s^a\f^a}&=&\vev{\gb^a\gb^a}=\k^4
\ee
the two classes of models  have {\it exactly} the same transverse
gluon polarization at one loop. To verify this, it is sufficient to
observe that the  relation\equ{vacuumrel} between the expectation
values of the moduli-fields of the 
two classes of models implies that $\vec\s$ and $\vec\f$ are
anti-parallel and that the length of $\gb$ is the geometric mean of
the lengths of $\s$ and $\f$. For such  configurations  the
difference between the  gluon polarizations in the two gauges is only
in  the  longitudinal parts  of $\G^{2\,ab}_{\mu\nu}(q; c))$
in\equ{contributions1}  and  $\G^{2\,ab}_{\mu\nu}(q; d))$
in\equ{contributions2}. 

The corrections to the transverse gluon polarization affect physical
correlation functions such as $\vev{ F^2(x) F^2(0)}$. We thus
believe that\equ{vacuumrel} holds in lowest order, although
this can only be verified directly by nonperturbative methods or as
the consequence of a Ward identity.  The different power corrections
to the {\it longitudinal}  gluon polarization are expected due to the
intrinsic gauge dependence of this object. They have no physical
consequences at this order of perturbation theory.  The
generalization of these remarks to higher (or all) orders in the loop
expansion is clearly desirable but beyond the scope of this paper.

\section{Conclusion }
We have reinterpreted the Faddeev-Popov procedure as the
construction of a TQFT in the gauge group depending on a background
connection. To handle the global Faddeev-Popov zero modes in covariant
Landau type gauges we considered an equivariant cohomology. The
elimination of the constant  zero modes is particularly
important at finite space-time volume, since they otherwise lead to a
vanishing partition function of the Yang-Mills theory. Contrary  to
pointed gauges~\cite{si78,mi81}, our proposal preserves the
covariance and translational invariance of gauge dependent Green
functions. We believe this to be relevant for 
defining the infinite volume limit of an unbroken nonabelian
gauge theory. It is otherwise not certain that
translational invariance of gauge dependent correlators can be
recovered in the infinite volume limit, since the color
forces of such theories are expected to be  strong and  of infinite range.

To implement covariant constraints and build the equivariant
cohomology we introduced constant ghosts. Most of them can be
eliminated from the effective action by equations of motion and the
remaining ones may be interpreted as moduli. The topological observables of the
gauge group TQFT are elements of descent equations that terminate in
operators which are only functions of the constant ghost $\f$ which
controls the translational symmetry of the Faddeev-Popov ghost. The winding
number of gauge group elements is the only topological observable
with ghost number zero we can construct in this way (apart from
linked observables derived  from this one). It is associated with the
invariant $\tr\f^2$ through the descent equations. Using the
properties of this observable, we have been able to show in a very
simple way that the partition function of the gauge group TQFT (and
thus the contribution from 
the gauge-fixing to the Yang-Mills theory) either vanishes due to
generic zero modes or depends on 
global properties of the connection. The latter situation occurs for
space-time manifolds admitting 3-cycles 
$\G_3$ with $\pi_3(\G)=\integers$. This is precisely the condition
found previously by Singer~\cite{si78} to be associated with a topological
Gribov problem. 

For an $SU(2)$ gauge group we verified
that our improved gauge fixing yields a well-defined and non-vanishing
partition function in the vicinity of the trivial Yang-Mills connection. The
argument in Appendix~A relies on the observation that the partition
function of a TQFT calculates the Euler 
number of the space of fixed points in ``delta-'' (Landau-) gauge. 
By analyzing the topological structure of the space of fixed points
for $A=0$ we concluded that its Euler number is {\it odd}. The
partition function of our  equivariant TQFT   therefore
does not vanish generically. Together with\equ{dependence} this gives a direct
verification  of the dependence of the TQFT on global topological
properties in the vicinity of flat background connections.
We have shown that this global violation of the BRST symmetry can be
reformulated as the existence of a nonlocal but BRST-exact operator
whose expectation value does not vanish.

With the modified BRST algebra the usual
Feynmann rules are changed by couplings of the constant ghosts to the
ordinary Yang-Mills field content.  
We investigated the possible effects on perturbation theory and gauge
invariance by considering two classes of renormalisable gauges involving the
constant ghosts.  To  avoid  infrared problems in the perturbative
treatment of these gauge theories we reorganized the perturbative
expansion and treated constant bosonic ghosts like moduli.  Since there
is no such argument as a holomorphic dependence on the moduli as in
supersymmetric theories,  the true effective action for the moduli is
not known.  However, only a few expectation
values of the moduli have to be known in order to completely
determine the loop expansion of observables in the infinite
volume limit.  We computed one-loop effects and observed that  the two
classes of gauges  are physically equivalent provided one suitably
identifies expectation values of the constant ghosts. Nonvanishing expectation
values of the constant commuting ghosts lead to a modification of the
gluon polarization. At sufficiently high momenta, where a 
perturbative analysis is justified, we obtained power
corrections to the leading logarithmic behaviour. These corrections
are similar to those one  expects nonperturbatively for instance from
instantons, but are in our case the result of
interactions with the constant bosonic ghosts.  Moreover, we verified that
the power corrections to the 
transverse gluon polarization are gauge independent,
whereas the  modification of the longitudinal polarization is not.
This is a nontrivial consistency test for our equivariant construction,  but
it clearly would be of interest to prove the gauge independence of
physical observables beyond the one loop level that we investigated here.  

\vspace{.5cm}
\begin{center}
\bf Acknowledgements
\end{center}\nobreak
We would like to thank D. Zwanziger and P.~Mitter for stimulating
discussions and helpful comments. M.S. is also grateful for the
hospitality extended at the summer school of Carg\`ese and by the LPTHE.
We would also like to thank A.~Rozenberg and M.~Porrati for pointing
out an error in the original manuscript.

\begin{appendix}
\section{The partition function \ZZ[A=0]}
We show in this appendix that the partition function 
\eq
\ZZ[A]=\vev{\one}_A =\int [dU] [dc] [d\cb] [db] d\f d\s d\sb
d\gb  d\g\  e^{S_{A}} 
\eqn{norm}
of the equivariant TQFT with action\equ{actionA} no longer vanishes
generically for all connections $A$ and thus establish
that\equ{dependence} is non-trivial. 

The integrations over the constant ghosts
$\f$ and $\s$ in\equ{norm} can be performed, giving a
field-independent factor proportional to the inverse volume of space-time
(which we take to be finite here). The integration over the other fields is
more subtle, since we know that there are (infinitely) many configurations
$U(x)$ satisfying the saddle-point equation 
\eq
\pa\ipr A^U= \pa\ipr(U^\dagger A U + U^\dagger \pa U)=0\ 
\eqn{classical}
For $A=0$ these are the gauge-copies of the vacuum in Landau gauge
which were first considered by Gribov~\cite{gr78}  who also
constructed some of them. 
The solutions $U$ of\equ{classical} fall into equivalence classes modulo
constant (right) gauge transformations and the equivariant construction
is designed to handle this (trivial) degeneracy. 
A semiclassical evaluation of\equ{norm} means that one  
integrates in the vicinity of a representative for each class. This is  a 
difficult task, considering that there are arbitrary many such
copies. However, the contribution of a {\it single} copy to the
normalization is expected to be finite due to supersymmetric
compensation of the mode expansion. If there were no zero modes, the
contribution of {\it each} copy to\equ{norm} could be normalized to $\pm
1$, and the result of\equ{norm} would then be interpreted as the degree of
the map of the gauge-fixing~\cite{hi79}.

In the more general setting with (additional) zero modes, our 
equivariant partition function should compute the 
(generalized) Euler number $\chi(\EE_A/SU(n))$ of the topological
space of fixed points of\equ{classical} 
\eq
\EE_A:=\{U: U(x)\in SU(n), \pa\ipr A^U=0\}
\eqn{space} 
rather than the Euler number of $\EE_A$ itself~\cite{bi91}. We will
argue that $\chi(\EE_{A=0}/SU(2))=odd\neq 0$ for an $SU(2)$ gauge group. The
equivariant partition function\equ{norm} therefore should not vanish
in the vicinity\footnote{It is conceivable that the
(generalized) Euler number the partition function computes may depend
on topological characteristics of the background connection
$A$. Instructive in this respect is that $\EE_A$ can be viewed as the
space of fixed points of an associated Morse potential~\cite{zw94}
$\VV_A[U]\in \RR_+$ 
\eq
\VV_A[U] = \int_\MM \tr A^U\ipr A^U
\eqn{Morse}
Taken naively, the Morse theorem would indicate that the TQFT is
actually calculating the (generalized) Euler number 
of $\{U: U(x)\in SU(n), V_A[U]<\infty\}/SU(n)$. The
Morse potential\equ{Morse} thus resticts the functional space of  gauge
transformations one is considering in an $A$-dependent fashion. This
statement should however be taken with a grain of salt, since it is far from
obvious that Morse theory is applicable in this infinite-dimensional
setting.} of $A=0$. 
The action\equ{actionA} at $A=0$ is also
symmetric with respect to {\it left} multiplication by a (constant) group
element $U(x)\rightarrow g_L U(x)$. Consequently, if $U(x)$ is a
solution to\equ{classical} at $A=0$, 
\eq
\pa\ipr(U^\dagger \pa U)=0
\eqn{classical0}
then so is
\eq
U(x,g_L,g_R) = g_L U(x) g_R,  \qquad \forall g_L,\,g_R\in SU(n)
\eqn{mods}
Some of these solutions however belong to the same equivalence class
modulo right multiplication by $SU_R(n)$. We obtain the
moduli space of these (right equivalence classes of) solutions
to\equ{classical0} by noting that left
multiplication of $U(x)$ by $g_L\in SU(n)$ is the same as right
multiplication by  
\eq
g_R(x)=U^\dagger(x) g_L U(x)
\eqn{eqgr}
$g_L U(x)$ therefore belongs to the equivalence class modulo right
multiplication of $U(x)$ only if $d g_R(x)=0$, i.e.
\eq
[U(x)d U^\dagger(x),\, g_L]=0
\eqn{equivlr}
Thus left multiplication of $U(x)$ by any $g_L\in SU(n)$ belonging
to the subgroup which commutes with $U(x) d U^\dagger(x)$ is
redundant. For an $SU(2)$ gauge group there are  only three possible
subgroups to consider: 
\begin{itemize}
\item[1)] $g_L\in  SU_L(2)$ satisfy\equ{equivlr} $\Rightarrow
 \chi(SU_L(2)/SU_L(2))=1$ 
\item[2)] $g_L\in U(1)\subset SU_L(2)$ satisfy\equ{equivlr} $\Rightarrow
\chi(SU_L(2)/U(1)\simeq S_2)=2$
\item[3)] $g_L\in \{\one, -\one\} \subset
SU_L(2)$ satisfy\equ{equivlr} $\Rightarrow
\chi(SU_L(2)/\{\one,-\one\}\simeq SO(3))=0$    
\end{itemize}
Note that case 1) implies that $U(x) d U^\dagger(x)=0$ and therefore
corresponds to the equivalence class of the identity (modulo right
multiplication). There is only  {\it one} such class. For an $SU(2)$
gauge group, case 2) can only occur if $U(x)d U^\dagger(x)=
d\t(x)$ is an {\it abelian} connection. Since $U(x)$  should
furthermore satisfy\equ{classical0}, we can conclude that this
connection vanishes unless there are non-trivial 1-cycles.
We thus find that the moduli space
of solutions to\equ{classical0} of an $SU(2)$ theory has the
topological structure
\eq
\EE_{A=0}=SU_R(2)\times\left[\one + S_2\times\FF+ SO(3)\times
\widetilde\FF\right]  
\eqn{structure}
In other words $\EE_{A=0}/SU(2)$ can essentially be described as a
single point and a collection of two- and three-dimensional
spheres~\cite{si78, vb95}. 
Although we do not know the Euler numbers associated with the
topological spaces  $\FF$ and $\widetilde\FF$, the
structure\equ{structure}, together with $\chi(S_2)=2$ and
$\chi(SO(3))=0$ suffices to see that  
\eq
\chi(\EE_{A=0}/SU(2))= odd\neq 0
\eqn{espace}
In the vicinity of $A=0$ the degeneracy with respect to left group
multiplication is in general lifted, but the signed sum of Morse
indices over the (then isolated) fixed points should still give the Euler
number\equ{espace}. At least for an $SU(2)$ gauge group, the partition
function\equ{norm} of the TQFT should therefore not vanish near
$A=0$.  We thus have  explicitly verified \equ{dependence} for the
$SU(2)$ case. 
\end{appendix}

}\end{document}